\documentclass[%
superscriptaddress,
 amsmath,amssymb,
 aps,
prb,
twocolumn
]{revtex4-1}

\usepackage{color}

\usepackage{graphicx}
\usepackage{dcolumn}
\usepackage{bm}
\usepackage{hyperref}

\begin{document}
\newcommand{\fig}[1]{Fig.~\ref{#1}}
\newcommand{\eq}[1]{eq.~(\ref{#1})}
\newcommand{\etal}{\textit{et al.}}

\preprint{APS/123-QED}

\title{ Properties of hydrogen, helium, and silicon dioxide mixtures in giant planet interiors}

\author{Fran\c{c}ois Soubiran}
 \affiliation{Department of Earth and Planetary Science, University of California, Berkeley, CA 94720, USA}
\author{Burkhard Militzer}
 \affiliation{Department of Earth and Planetary Science, University of California, Berkeley, CA 94720, USA}
 \affiliation{Department of Astronomy, University of California, Berkeley, CA 94720, USA}
\author{Kevin P. Driver}
 \affiliation{Department of Earth and Planetary Science, University of California, Berkeley, CA 94720, USA}
\author{Shuai Zhang}
 \affiliation{Department of Earth and Planetary Science, University of California, Berkeley, CA 94720, USA}

\date{\today}

\begin{abstract}
Recent observations of Jupiter and Saturn provided by spacecraft missions, such as 
Juno and Cassini, compel us to revise and improve our models of giant planet 
interiors. Even though hydrogen and helium are by far the dominant species in 
these planets, heavy elements can play a significant role in the structure and 
evolution of the planet. For instance, giant-planet cores may be eroded by their 
surrounding fluid, which would result in a significantly increased concentration of 
heavy elements in the hydrogen-helium envelope. Furthermore, the heavy elements 
could  inhibit  convection by creating a stabilizing gradient of composition. In 
order to explore the effects of core erosion, we  performed \textit{ab initio} 
simulations to study structural, diffusion and viscosity properties of dense 
multi-component mixtures of hydrogen, helium, and silicon dioxide at relevant 
pressure-temperature conditions. We computed radial distribution functions to 
identify changes in the chemical behavior of the mixture and to reveal 
dissociation trends with pressure and temperature. The computed diffusion 
coefficients of the different species as well as the viscosity provide constraints 
for the time scale of the dynamics of the core erosion and the mixing of its 
constituents into the envelope, which will help improve planetary models.
\end{abstract}

\pacs{Valid PACS appear here}

\maketitle

\section{\label{sec:intro}Introduction}
Gas giant planets, such as Jupiter and Saturn, are predominantly comprised of 
hydrogen and helium, existing in various phases. The core accretion model for 
giant planet formation require the accretion of heavy rocky or icy core materials 
before the planets can further accrete gas \cite{Pollack1996a}. Even after gas 
have been accreted, the planet also continues to capture planetesimals, but their 
fate remains unclear as they may dissolve within the gaseous envelope or reach the 
core of the planet\cite{Fortney2013}. However, the present-day distribution of 
these heavy elements in gas giants is not completely understood. Furthermore, the 
planet may not be completely differentiated into a core and an envelope. 
\textit{Ab initio} Gibbs free energy calculations using thermodynamic integration 
(TDI) showed that all the main species -- H$_2$O, SiO$_2$, MgO and Fe -- that 
could be found in the core are miscible in the metallic hydrogen that composes the 
deep region of the envelope and engulfs the core 
\cite{Wilson2012a,Wilson2012b,Wahl2013,Gonzalez-Cataldo2014}. Thus, the core could 
be progressively eroded by the surrounding H-He envelope but the time scale of 
this erosion process remains uncertain. If the time scale is short, then  the 
cores of Jupiter and Saturn would most likely be completely dissolved today, 
resulting into a heavy-element-enriched H-He fluid envelope. In contrast, if the 
erosion time scale is long compared to the age of the planet, the core remains 
mostly intact, resulting in a H-He envelope of near solar composition.

Except for oxygen, the Galileo entry probe measured a heavy element concentration 
(metallicity) in the outer envelope of Jupiter that is three times higher than 
solar metallicity \cite{Wong2004}, which is an indication that heavy elements are 
mixed in with the H-He fluid. The Cassini and Juno missions will measure the 
gravitational fields of Saturn and Jupiter with high precision, which will help to 
constrain the mass distribution in the planet. Combining insights from such data 
with improved planetary models, it is possible to infer the distribution of the 
heavy elements and to constrain the size of a core 
\cite{Nettelmann2012,Helled2013,Militzer2013b,Hubbard2016,Militzer2016}. An 
important step forward has already been achieved by computing more accurate 
equations of state (EOS) for the relevant constituents. We have performed 
\textit{ab initio} TDI calculations to determine the EOS of H-He mixtures 
\cite{Militzer2013a} and we investigated the influence of heavy materials on the 
thermodynamic properties of the envelope by performing simulations of mixtures of 
H, He and heavy elements \cite{Soubiran2016}.

The redistribution of heavy elements of the core into the envelope may also 
strongly depend on the dynamics of the fluid right on top of the core. Even if the 
erosion were fast, the mixing of the erosion products into layers directly above 
the core may be a very slow and difficult process due to effects of gravity. 
Convective forces may be too weak to  dredge up  heavy materials  into the outer 
layers of the planet \cite{Stevenson1982}. It is thus possible that a 
semi-convective or even stratified layer could occur above the core, significantly 
slowing down the mixing of the heavy elements with the envelope 
\cite{Leconte2012,Leconte2013}. The behavior of the heavy-element-bearing fluid 
can be readily determined by characterizing its transport properties. The 
transport properties of hydrogen and helium mixtures without heavy elements have 
already been partially investigated numerically \cite{French2012}, but the effect 
of the inclusion of heavy elements needs to be addressed.

In this article, we report results from \textit{ab initio} simulations of 
multi-component mixtures of hydrogen, helium and silicone dioxide. We focus on the 
dilute limit of a few percent in number of SiO$_2$. We explore the microscopic 
structure of the system using pair-correlation functions in order to identify 
chemical bonds. We determine the diffusion coefficients of the different species 
and the viscosity of the fluid. These quantities are crucial for determining the 
dynamics near the core-envelope boundary.

\section{\label{sec:methods}Methods}
\subsection{Simulation methods}
The results presented in this article rely on a series of molecular dynamics (MD) 
simulations based on density functional theory (DFT), using the Vienna \textit{ab 
initio} simulation package \cite{Kresse1996}. We set up the simulation in a cubic 
cell with periodic boundary conditions encompassing 220 hydrogen and 18 helium 
atoms following Militzer's set up \cite{Militzer2013a} and added from 2 to 4 
SiO$_2$ entities. We used a 0.2~fs time step for a total duration of a minimum of 
1~ps. The simulations were performed at constant volume and constant temperature 
using a Nos\'e thermostat \cite{Nose1984, Nose1991}. 

At each time step a DFT calculation was performed to determine the electronic 
density. We used the Kohn-Sham scheme \cite{Kohn1965} at finite temperature 
\cite{Mermin1965} and populated the eigenstates using a Fermi-Dirac distribution. 
Following previous work on H-He mixtures \cite{Militzer2013a}, we used the Perdew, 
Burke and Ernzerhof (PBE) exchange correlation functional \cite{Perdew1996} which 
uses the generalized gradient approximation (GGA). To improve the efficiency of 
the calculation, we used projector augmented wave (PAW) pseudo-potentials 
\cite{Blochl1994} including a frozen core for oxygen and silicon atoms only. We 
used a 1200~eV energy cutoff for the plane-wave basis set. The number of bands was 
adjusted to the concentration, density and temperature so that the spectrum of 
fully and partially occupied eigenstates was fully covered. We used the 
Baldereschi point \cite{Baldereschi1973} to sample the Brillouin zone, which was 
found to be sufficient in the case of H-He mixtures \cite{Militzer2013a}. We 
investigated the properties of the mixtures from 5000~K to 15000~K and from 25~GPa 
to 2000~GPa covering the metallic region of the interior of gas giant planets 
nearly up to the core boundary.

\subsection{\label{ssec:autocorr}Calculation of the ionic transport properties}
From the DFT-MD trajectory, it is possible to extract information on the dynamics 
of the system. For instance, we can use the fluctuation-dissipation theorem and 
its applications on the autocorrelation functions to determine the ionic transport 
properties of the mixtures. The autocorrelation function of the velocity is 
related to the diffusion coefficient of species $\alpha$ by a Green-Kubo formula 
\cite{Frenkel2002,Danel2012},
\begin{equation}\label{eq:defdiff}
D_\alpha=\frac{1}{3N_\alpha}\sum_i^{N_\alpha}\int_0^{+\infty}\langle\textbf{v}
_i(\tau)\cdot\textbf{v}_i(0)\rangle \;\textrm{d}\tau,
\end{equation}
where $N_\alpha$ is the number of particles of type $\alpha$ and 
$\textbf{v}_i(\tau)$ is the velocity vector of the $i$th particle at time $\tau$. 
The brackets represent an ensemble average over multiple origins along the full 
length of the simulation \cite{Allen1987}.  

Similarly, it is possible to determine the viscosity by computing the 
autocorrelation function of the off-diagonal components of the stress-tensor: 
\begin{equation}\label{eq:defvisc}
\eta=\frac{V}{3k_\textrm{B}T}\sum_{\{ij\}}\int_0^{+\infty}\langle\sigma_{
ij}(\tau)\sigma_{ij}(0)\rangle \;\textrm{d}\tau,
\end{equation}
where $V$ and $T$ are the volume and the temperature respectively, $k_\textrm{B}$ 
the Boltzmann constant and $\sigma_{ij}(\tau)$ the $ij$ off-diagonal component of 
the stress-tensor, with $ij$ in $\{xy,yz,xz\}$ in Cartesian coordinates. 

Because of the small number of heavy element atoms and the finite length of the 
simulations, the autocorrelation functions and thus the integrals can become very 
noisy. In order to minimize the effect of the noise , we followed Meyer \textit{et 
al.}'s methodology \cite{Meyer2014} and fitted the autocorrelation function with a 
multi-time scale function. For the velocity autocorrelation function 
$\gamma_\textbf{v}(t)=\frac{1}{3N_\alpha}\sum_i^{N_\alpha}\langle\textbf{v}
_i(t)\cdot\textbf{v}_i(0)\rangle$, we used the following expression:
\begin{equation}\label{eq:fitVACF}
\gamma_\textbf{v}(t)=a_0e^{-t/\tau_0}+\sum_{l=1}^{l_\textrm{max}}e^{-t/\tau_l}
\left[\mu_l\cos(\omega_l t)+\nu_l\sin(\omega_l t)\right]
\end{equation}
The parameters $a_0$, $\tau_0$, $\{\mu_l, \nu_l,\tau_l,\omega_l\}_l$ were 
determined with a least-square fit. We used parameters up to $l_\textrm{max}=1$ 
for H and He. For Si and O we needed to go up to $l_\textrm{max}=3$, but, for some 
cases, the multi-time scale fit did not converged and we had to employ the 
exponential decay only (first term in Eq. \ref{eq:fitVACF}). Once the parameters 
have been determined, the integration of Eq. \ref{eq:defdiff} yields:
\begin{equation}\label{eq:diffafterfit}
 D_\alpha=a_0\tau_0+\sum_l \tau_l \frac{\mu_l+\nu_l \omega_l \tau_l}{1+\omega_l^2 
\tau_l^2}
\end{equation}
For the autocorrelation function of the stress tensor 
$\gamma_\sigma(t)=\frac{1}{3}\sum_{\{ij\}}\langle\sigma_{ij}
(t)\sigma_{ij}(0)\rangle$, we used a simpler functional form:
\begin{equation}\label{eq:stresstensorfit}
 \gamma_\sigma(t)=Ae^{-t/\tau_1}+Be^{-t/\tau_2}\sin(\omega t),
\end{equation}
where $A$, $B$, $\tau_1$, $\tau_2$ and $\omega$ are fit parameters. Inserting Eq. 
\ref{eq:stresstensorfit} into Eq. \ref{eq:defvisc}, we obtained the following 
expression for the viscosity,
\begin{equation}
 \frac{k_\textrm{B}T \eta}{V} = A\tau_1+\frac{B\tau_2}{1+\omega^2\tau_2^2}.
\end{equation}

\begin{figure}[!ht]
\centering
\includegraphics[width=\columnwidth]{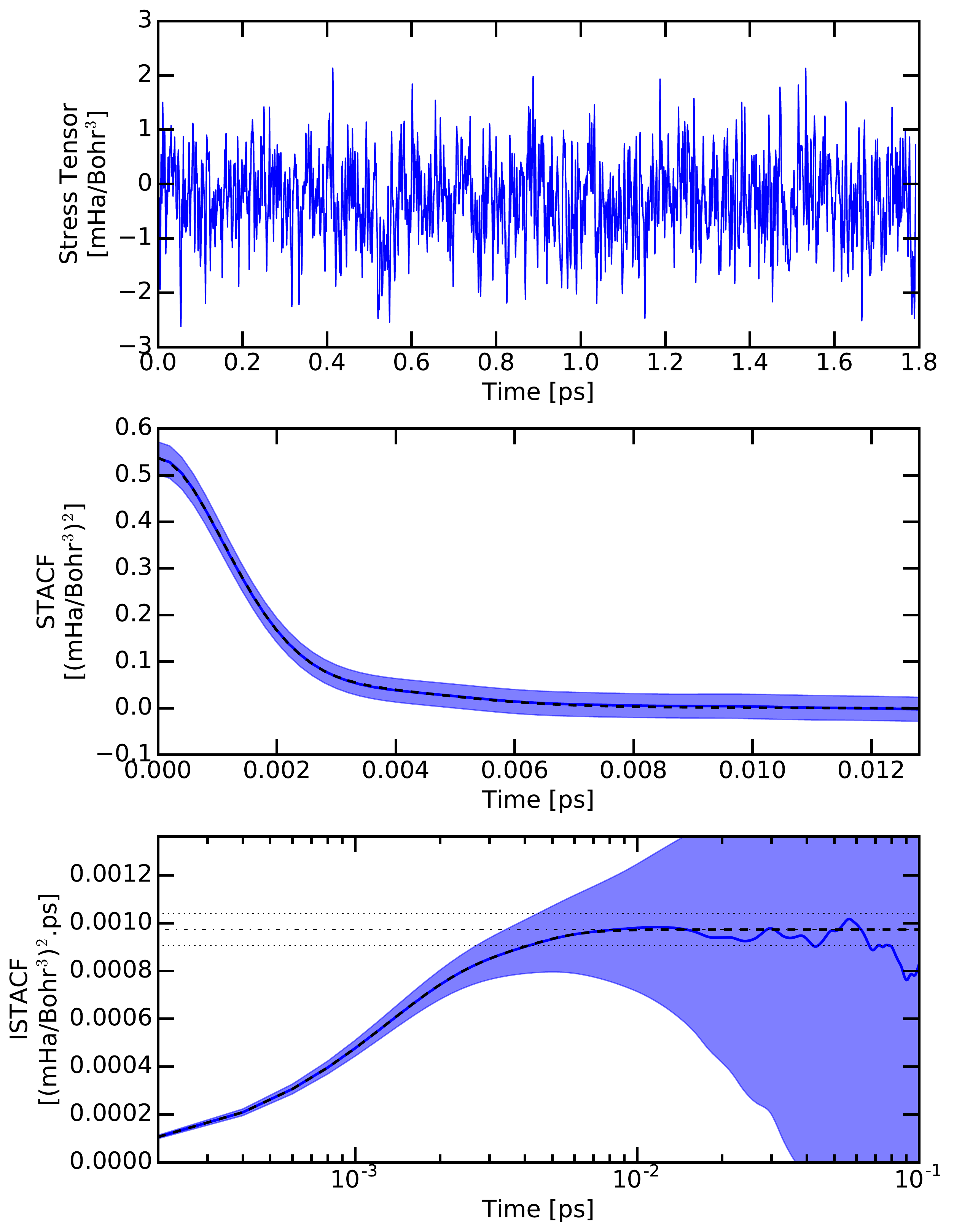}
\caption{\label{fig:explSTACF} (Color online) Example of a calculation of the 
viscosity of a 220 H, 18 He, 3 SiO$_2$ mixture at 15000~K and 1600~GPa. The top 
panel shows the xy-component of the stress-tensor. The middle panel shows in blue 
the autocorrelation function of the stress-tensor (STACF) and its uncertainty. The 
black dashed line is the fit with formula (\ref{eq:stresstensorfit}). The bottom 
panel shows the integral of the STACF and of the fit. The dash-dotted-line shows 
the asymptote of the fit and its statistical one-sigma uncertainty with the dotted 
line.  }
\end{figure}

As an example, we show in \fig{fig:explSTACF} the fit of the stress-tensor 
autocorrelation function of a 220 H, 18 He, 3 SiO$_2$ mixture at 15000~K and 
1600~GPa. The agreement between the fit and the actual autocorrelation function is 
excellent giving confidence on the value of the viscosity despite the high level 
of noise on the integral of the autocorrelation function also plotted in 
\fig{fig:explSTACF}. The use of the multi-time scale allow us a much more closer 
fit of the autocorrelation function.

\section{\label{sec:results}Results and discussion}
\subsection{\label{ssec:rdf} Structure of the mixture}
The simulations we performed cover a large range of conditions and different 
behaviors are expected for the mixture. For instance, at low pressure and 
temperature, hydrogen is molecular but it dissociates and becomes metallic when 
the temperature or the pressure increases \cite{Collins2001,Morales2010}. It has 
been shown that adding helium stabilized the molecular phase near the 
molecular-to-metallic transition \cite{Vorberger2007}. In order to investigate the 
influence of adding SiO$_2$ to the system, we computed the pair-correlation 
function between particles of type $\alpha$ and $\beta$:
\begin{equation}
g_{\alpha\beta}(r)= \frac{V}{4\pi r^2N_\alpha 
N_\beta}\left\langle\sum_{i_\alpha,j_\beta}\delta(r
-|\textbf{r}_{i_\alpha}-\textbf{r}_{j_\beta}|)\right\rangle,
\end{equation}
where $r$ is a variable that typically varies from 0 to half the size of the 
simulation box, $\textbf{r}_{i_\alpha}$ is the position vector of the 
${i_\alpha}$th particle of type $\alpha$ and the brackets denote a time average 
over the duration of the MD simulation. The pair-correlation function is a 
standard tool to characterize the microscopic arrangement around each type of 
particle and can help determine the existence of molecular bonds 
\cite{Soubiran2015a,Soubiran2015}.

\begin{figure}[!ht]
\centering
\includegraphics[width=\columnwidth]{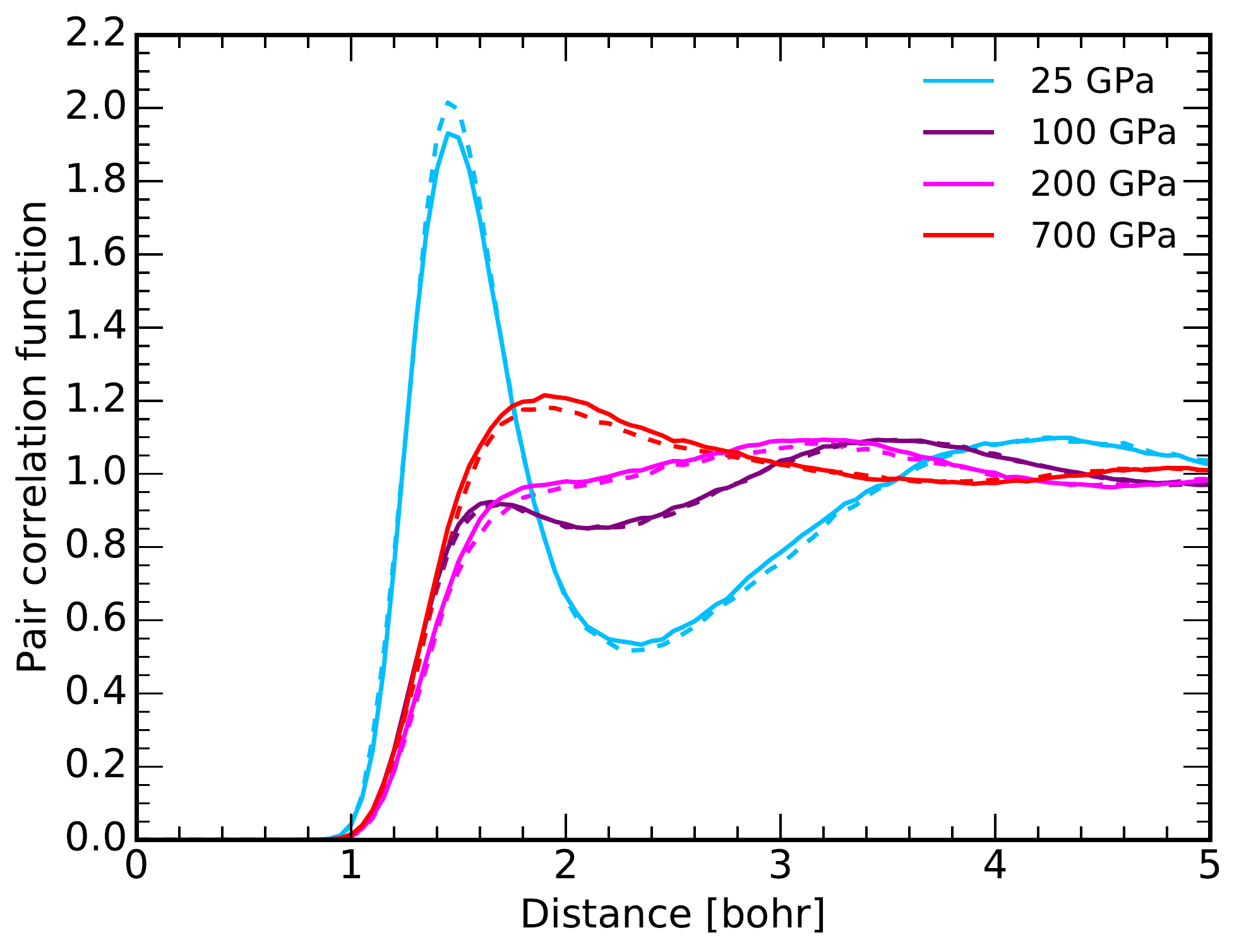}
\caption{\label{fig:rdfHH5000} (Color online) H-H pair-correlation function in 
mixtures with 220 H, 18 He and 2 (resp. 4) SiO$_2$ in dashed (resp. full) line. 
The temperature is at 5000 K and the pressure ranges from 25 to 700~GPa as 
indicated in the legend. }
\end{figure}

\begin{figure}[!ht]
\centering
\includegraphics[width=\columnwidth]{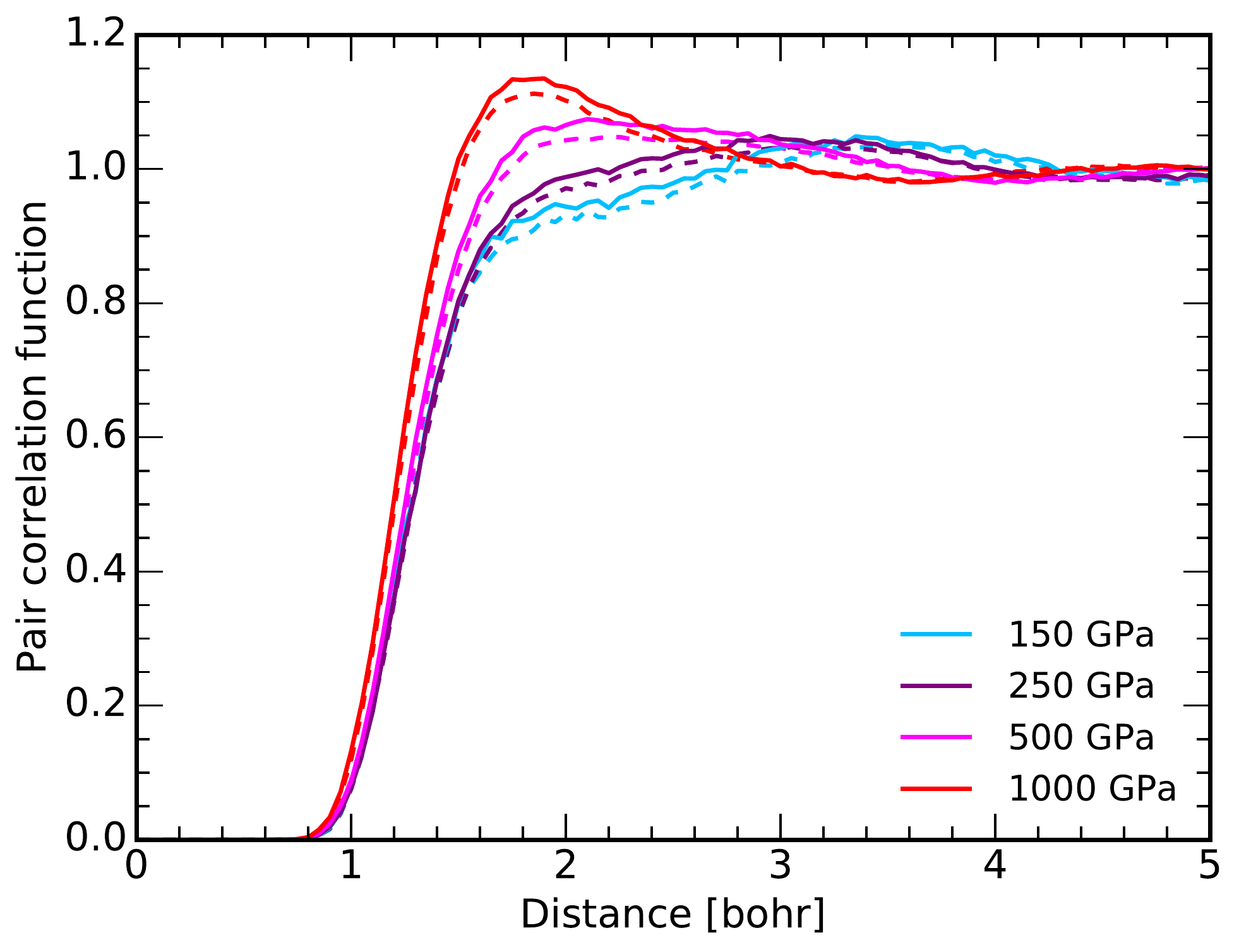}
\caption{\label{fig:rdfHH10000} (Color online) H-H pair-correlation function in 
mixtures with 220 H, 18 He and 2 (resp. 4) SiO$_2$ in dashed (resp. full) line. 
The temperature is at 10000 K and the pressure ranges from 150 to 1000~GPa as 
indicated in the legend. }
\end{figure}

\fig{fig:rdfHH5000} shows that the pair-correlation function between hydrogen 
atoms changes significantly as the pressure increases for a fixed temperature of 
5000~K. Over the entire pressure range, there is an exclusion radius of about 
1~bohr. At 25~GPa, the sharp peak at 1.4~bohr represents H$_2$ molecules meaning 
that the system is not entirely dissociated. But this peak disappears as the 
pressure increases indicating a complete dissociation of the molecules. This is 
consistent with observations in pure hydrogen \cite{Caillabet2011} and in H-He 
mixtures \cite{Vorberger2007}. It is very interesting to note that the peak height 
slightly decreases as more SiO$_2$ entities are inserted. Unlike helium, SiO$_2$ 
does not stabilize the H$_2$ molecules. At 10000~K, as plotted in 
\fig{fig:rdfHH10000} the 1.4~bohr peak is completely smoothed out, which means 
that hydrogen is fully dissociated. At both temperatures, we see that the second 
peak shifts towards smaller radii as pressure increases, which is 
simply an effect of compression and the increase in density.

\begin{figure}[!ht]
\centering
\includegraphics[width=\columnwidth]{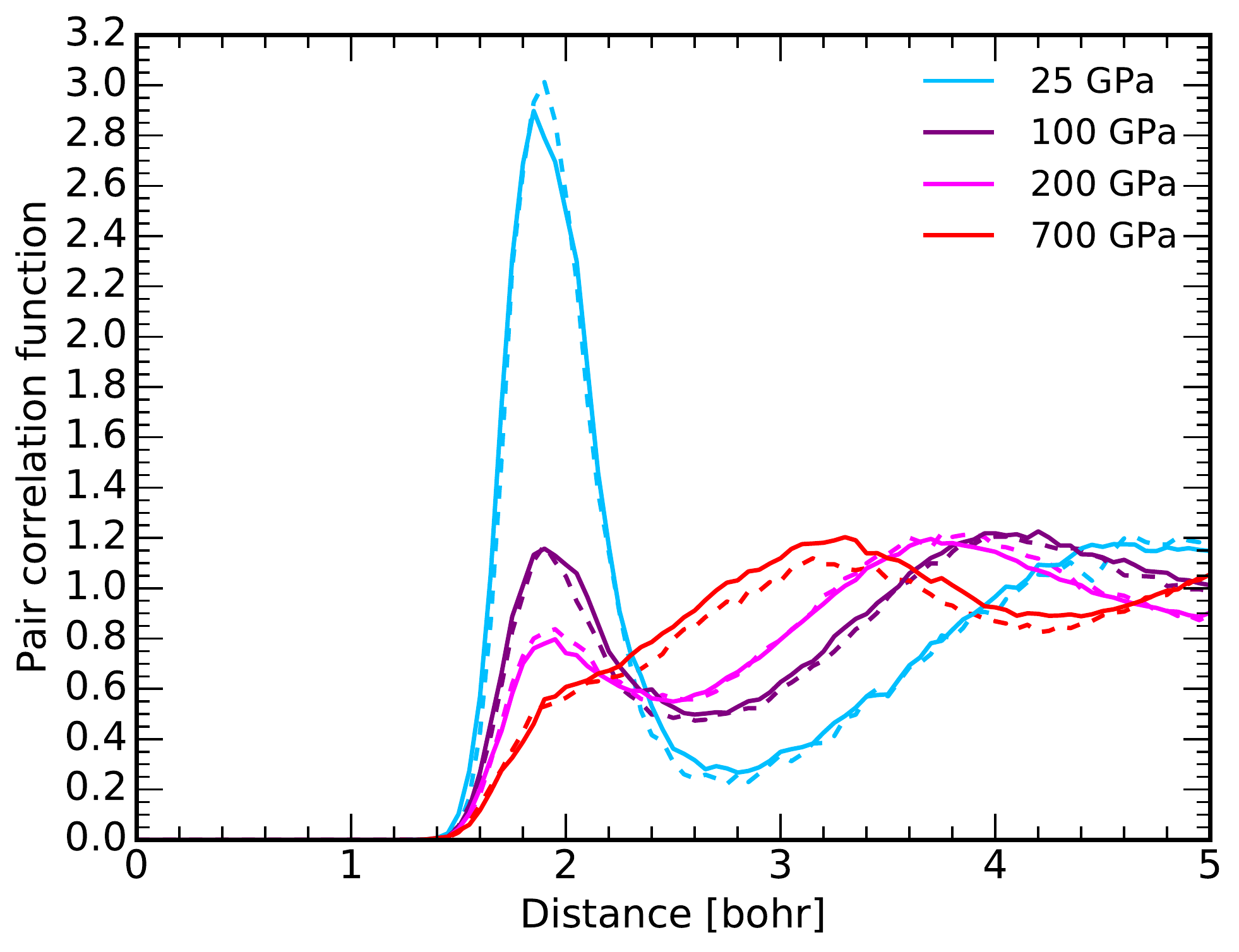}
\caption{\label{fig:rdfHO5000} (Color online) H-O pair-correlation function in 
mixtures with 220 H, 18 He and 2 (resp. 4) SiO$_2$ in dashed (resp. full) line. 
The temperature is at 5000 K and the pressure ranges from 25 to 700~GPa as 
indicated in the legend.}
\end{figure}

\begin{figure}[!ht]
\centering
\includegraphics[width=\columnwidth]{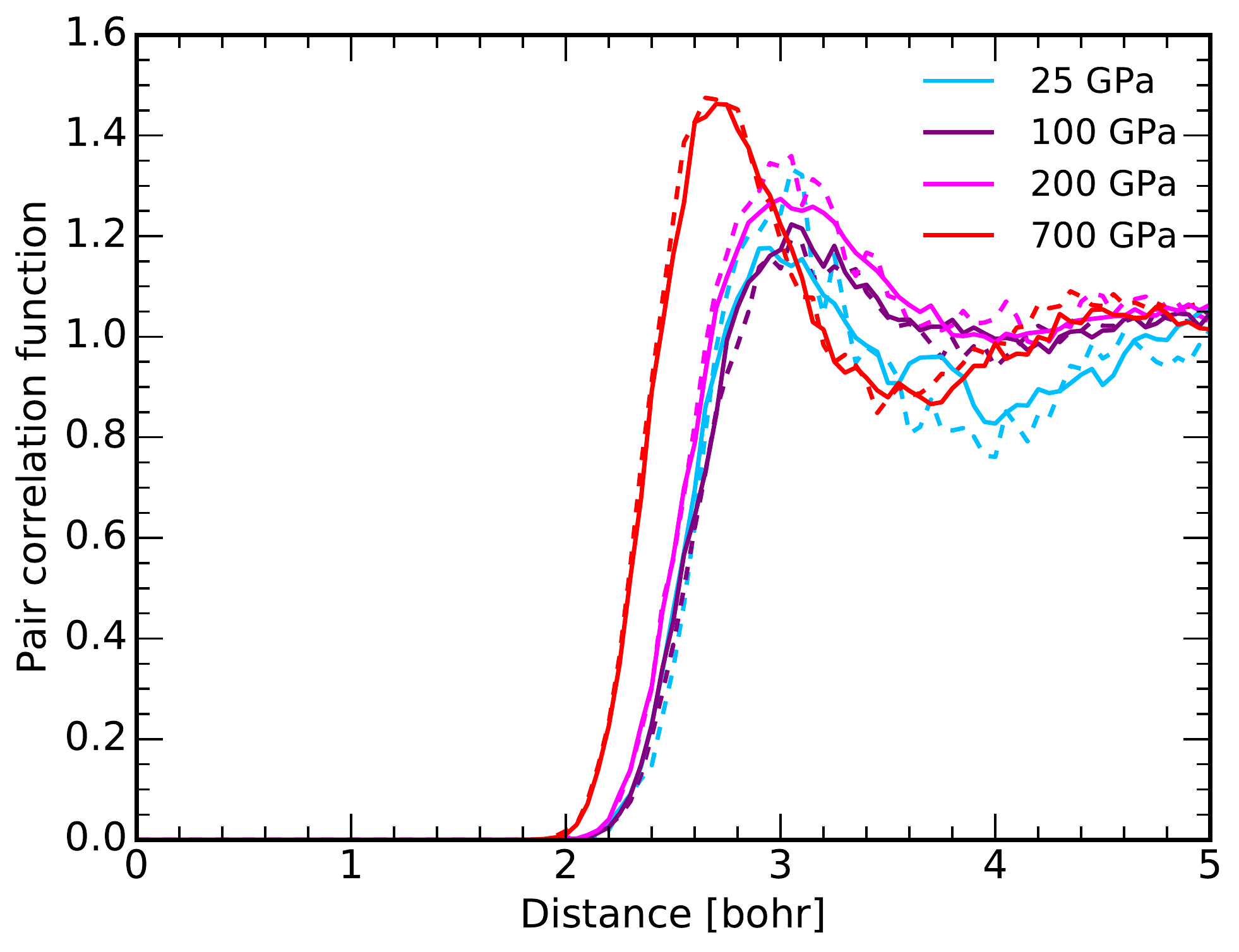}
\caption{\label{fig:rdfHSi5000} (Color online) H-Si pair-correlation function in 
mixtures with 220 H, 18 He and 2 (resp. 4) SiO$_2$ in dashed (resp. full) line. 
The temperature is at 5000 K and the pressure ranges from 25 to 700~GPa as 
indicated in the legend.}
\end{figure}

The O-H pair-correlation function, plotted in \fig{fig:rdfHO5000} for a 
temperature of 5000~K, shows similar features as the H-H pair-correlation 
function. We observe a sharp peak, around 2~bohr at 25~GPa, that progressively 
disappears as pressure increases. But this peak is visible longer than for H-H. 
This peak indicates that hydrogen bonds with oxygen to form hydroxide or water 
molecules. The existence of such molecules was confirmed by a cluster analysis of 
the simulations similar to the one we performed for hydrogen-water mixtures 
\cite{Soubiran2015}. At 25~GPa, we find that 45\% of the oxygen atoms are 
chemically bonded forming hydroxide molecules and 10\% are in water molecules. 
These molecules progressively dissociate but are slightly more stable than H$_2$ 
molecules, as was previously predicted for water-hydrogen mixtures 
\cite{Soubiran2015a,Soubiran2015}. The existence of H-O bonds also explains why 
the peak in the H-H pair-correlation function decreases with the number of 
inserted SiO$_2$. As the number of free oxygen atoms increases, the number of H-O 
bonds increases breaking H-H bondsbbecause the O-H bond is more stable than the 
H-H bond. This results in a dissociation of molecular hydrogen. The oxygen does 
not stabilize H$_2$ molecules, like helium. Instead, it chemically reacts with 
hydrogen. 

The H-Si pair-correlation functions at 5000~K, plotted in \fig{fig:rdfHSi5000}, do 
not exhibit the same molecular features. There is an exclusion radius of nearly 
2~bohr but the first peak does not exhibit the features of a molecular system. We 
can assume that hydrogen does not form stable bonds with silicon under these 
conditions. 

\subsection{\label{ssec:diff}Diffusion properties}
The dynamics of the core erosion in giant planets can be slightly impacted by the 
diffusion properties of the heavy elements. If the diffusion is fast, eroded 
materials can mix efficiently with the surrounding H-He fluid and the erosion is 
likely to be rapid. But if the diffusion is slow, it may render the erosion 
process extremely inefficient and thus there could still be a core in giant 
planets despite the thermodynamic predictions favoring the complete dissolution of 
the core into the envelope. 

To compute the diffusion properties, we calculated the autocorrelation function of 
the velocities as indicated in section \ref{ssec:autocorr}. We only report 
calculations for the highest concentration in SiO$_2$ because of the small number 
of heavy elements, which prevents us to have a good statistics. We do not expect 
however the diffusion coefficient to be dependent on the concentration of heavy 
elements at least in the diluted limit. 

\begin{figure}[!ht]
\centering
\includegraphics[width=\columnwidth]{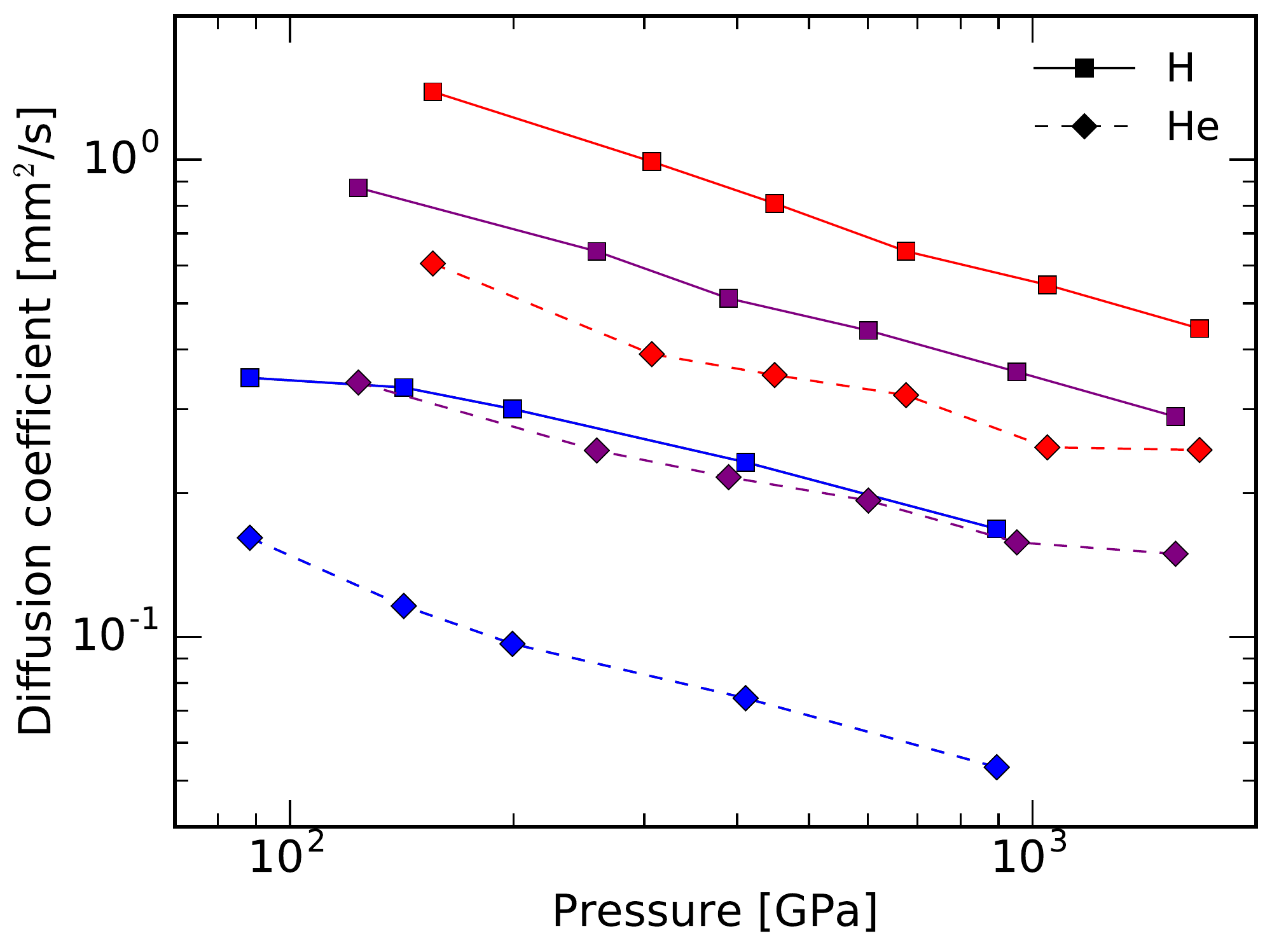}
\caption{\label{fig:diff_light} (Color online) Diffusion coefficients as a 
function of the pressure, for H and He in multi-component mixtures of 220 H, 18 He 
and 4 SiO$_2$, on three isotherms: 5000~K (blue), 10000~K (purple) and 15000~K 
(red).}
\end{figure}

The diffusion coefficients of H and He are plotted in \fig{fig:diff_light}. They 
show a very smooth evolution as a function of pressure and temperature. As the 
temperature increases, the diffusion is easier because the thermal velocity is 
higher. But the diffusion coefficients decreases as the pressure increases because 
the particles interact more strongly at higher density. It is also interesting to 
note that at 10000 and 15000~K we have roughly a factor of 2 difference between 
the diffusion coefficient of helium and that of hydrogen. But it increases to a 
factor of almost 4 at 5000~K. The calculated coefficients are in reasonable 
agreement with the diffusion coefficient for H-He mixtures as computed by French 
\textit{et al.} \cite{French2012}, indicating a marginal influence of the heavy 
elements on the diffusion of H and He. There is also a sort of plateau in the 
diffusion coefficient of hydrogen below 100~GPa at 5000~K. We observed a very 
similar feature in the diffusion of hydrogen in hydrogen water mixtures and 
attributed it to the dissociation of hydrogen and of water\cite{Soubiran2015}. It 
is safe to assume that the dissociation of H$_2$ is also at play in the present 
case. 

\begin{figure}[!ht]
\centering
\includegraphics[width=\columnwidth]{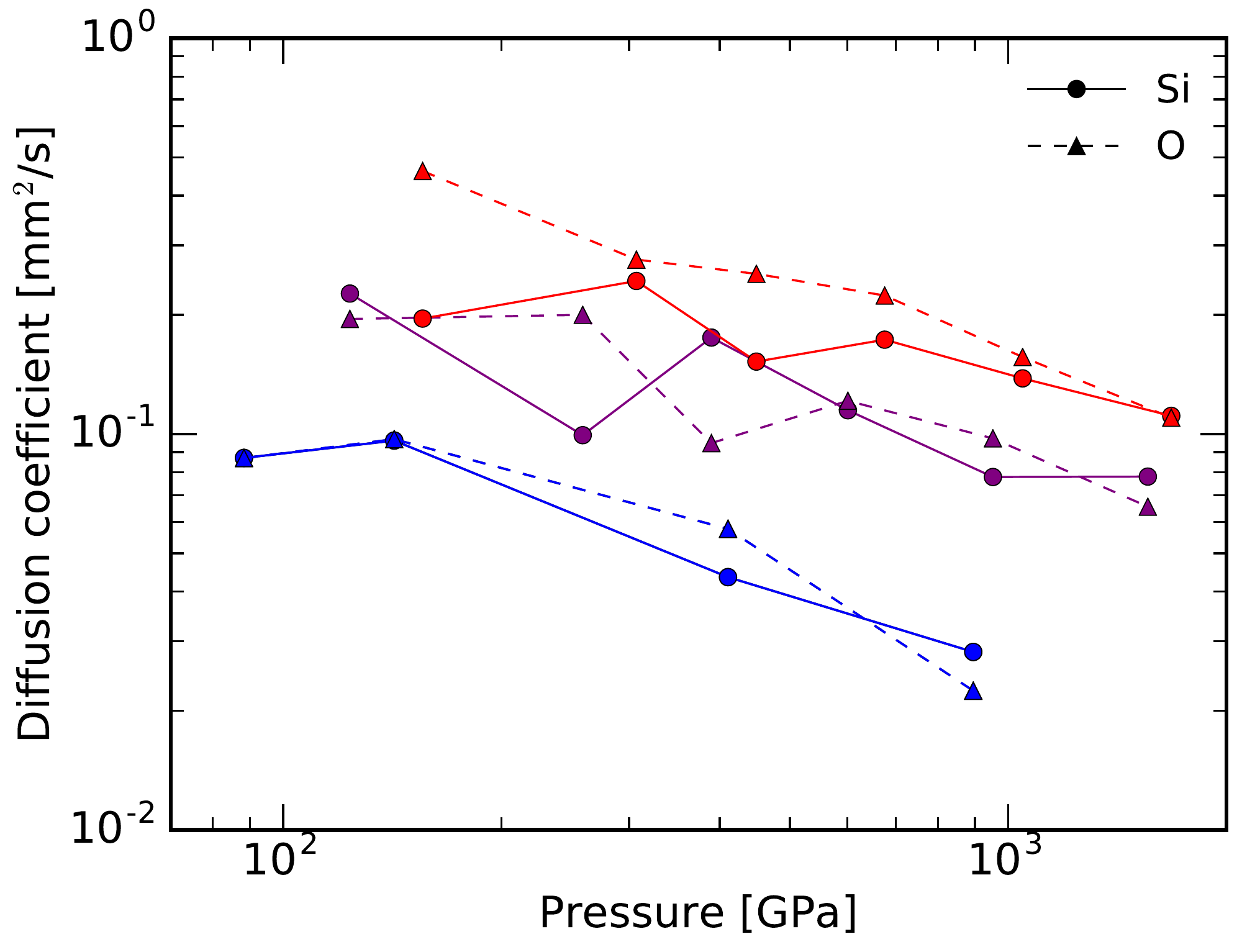}
\caption{\label{fig:diff_heavy} (Color online) Diffusion coefficients as a 
function of the pressure, for Si and O in multi-component mixtures of 220 H, 18 He 
and 4 SiO$_2$, on three isotherms: 5000~K (blue), 10000~K (purple) and 15000~K 
(red).}
\end{figure}

The calculations of the diffusion coefficients for the heavy elements is much more 
challenging because of the poor statistics due to the small number of heavy atoms 
but also because the decorrelation time is much longer as these particles move 
much more slowly, and we would need longer simulations to have a complete 
description of the autocorrelation function. Because of that, the fit is of lower 
quality and we were not always able to obtain a converged fit using the multi-time 
scale decomposition as in eq. (\ref{eq:fitVACF}). For some simulations, we had to 
use a simple exponential decay. Therefore, the diffusion coefficients presented in 
\fig{fig:diff_heavy} can only be considered as an order of magnitude.

The trends observed on H and He are also visible for the diffusion coefficients of 
Si and O despite the scatter in the curves in \fig{fig:diff_heavy}. The pressure 
has a negative effect on the diffusion while temperature makes the diffusive 
process easier. The oxygen and the silicon seem to have similar diffusion 
coefficients at low temperature but at 15000~K, oxygen diffuses faster than 
silicon as expected by the difference of mass. Overall, we observe a slower 
diffusion of the heavy elements compared to H and He as expected. However, the 
diffusion is only slower by one order of magnitude at most for the range of 
parameters we explored.

\subsection{\label{ssec:Viscosity}Viscosity}
It is usually assumed that the deep interior of giant planets is fully convecting 
\cite{Militzer2016} but the possible erosion of the core could inhibit the 
convection \cite{Leconte2012,Leconte2013} and a semi-convection pattern could set 
in instead. One key parameter in the stability analysis is the viscosity of the 
fluid. By computing the autocorrelation function of the off-diagonal components of 
the stress-tensor, we can determine the viscosity of the mixture as explained in 
section \ref{ssec:autocorr}.

\begin{figure}[!ht]
\centering
\includegraphics[width=\columnwidth]{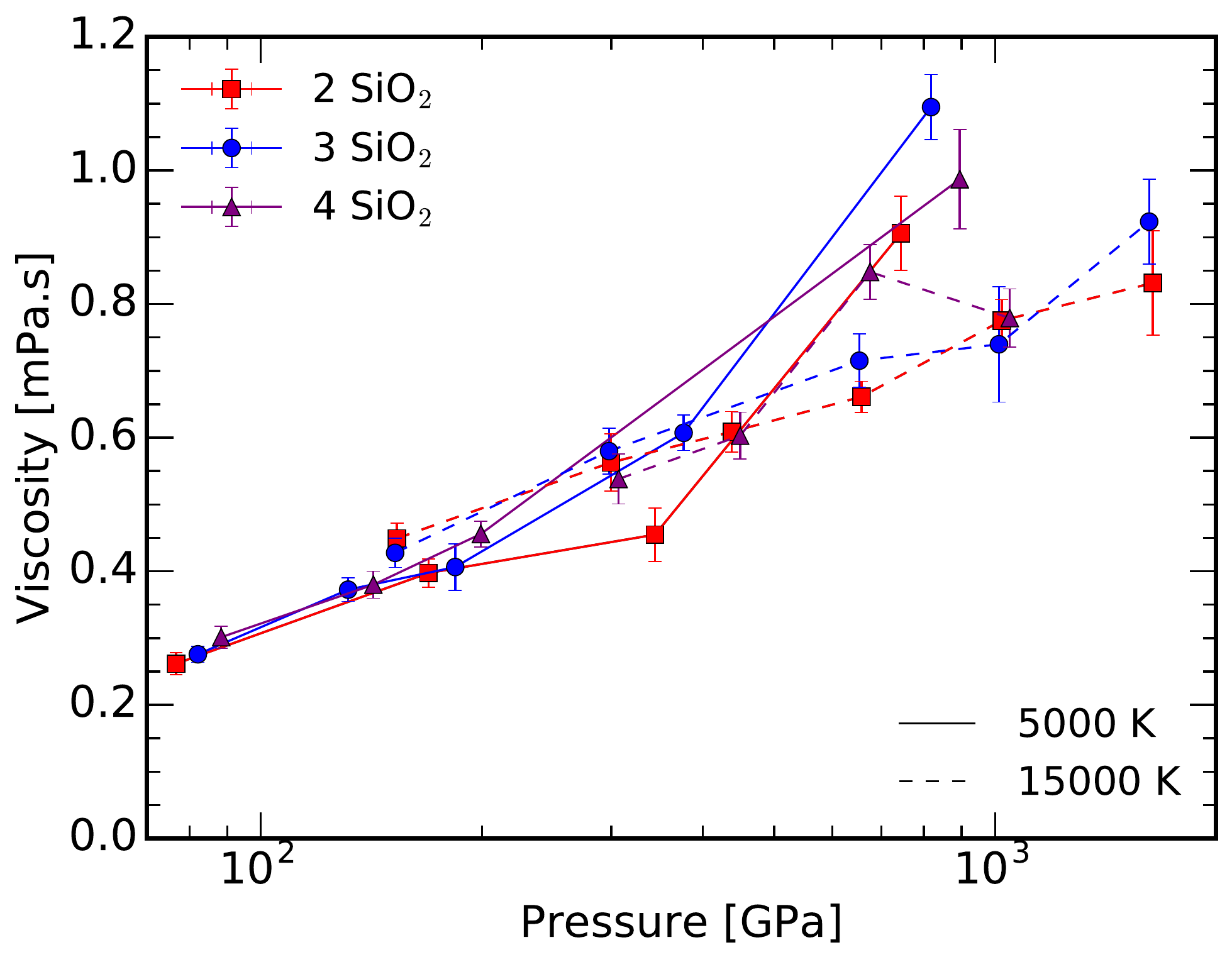}
\caption{\label{fig:viscosity} (Color online) Viscosity as a function of the 
pressure for mixtures with 220 H, 18 He and $N$ SiO$_2$ -- $N$ indicated in the 
legend -- along two isotherms at 5000 and 15000 K. The errorbars represent a 
one-sigma statistical uncertainty.}
\end{figure}

As shown in \fig{fig:viscosity}, the viscosity evolves smoothly as a function of 
the temperature and the pressure. The range of values goes from 10$^{-4}$ to 
10$^{-3}$~Pa.s, which indicates that the convection of the metallic phase in giant 
planets is most likely to be turbulent and not laminar. The expected Reynolds 
number is of the order of $Re\sim10^{12}$ for the convection in Jupiter 
\cite{Guillot2004}. The values that we obtained for the viscosity are in agreement 
with H-He mixture calculations in similar conditions \cite{French2012}. While the 
dependence in temperature is unclear at low pressure, at higher pressure we 
observe that the viscosity slightly decreases as the temperature increases but the 
difference is not significant. Because of the size of the errorbars and some 
uncertainty on the goodness of the fit of the stress-tensor autocorrelation 
function for every conditions investigated, we are unable to identify any 
influence from the addition of SiO$_2$. In the dilute limit, the properties of the 
multi-component mixture are mostly the same as that of the H-He mixture. The heavy 
elements barely influence the bulk viscosity 
of the system.

\subsection{\label{ssec:erosion}Erosion time-scale}
Based on the results we obtained, it is possible to do an estimate of a possible 
time scale for the erosion of the core in a Jupiter-like planet. If we neglect the 
time it takes for an atom to overcome the energy barrier when it separates from 
the solid phase to go to the surrounding fluid, we can estimate the erosion 
time-scale as the time necessary for the atom to diffuse through the boundary 
layer of the convective cell on top of the core. Indeed, a typical convective cell 
is homogeneously mixed except in its viscous boundary layers. In these layers, 
most of the mixing is performed through diffusion. By estimating the diffusion 
time of the heavy elements through the boundary layer, we can determine the time 
it takes for the atom to reach region of the fluids where it is actually 
well-mixed and thus how fast the core can dissolve into the envelope. Numerical 
simulations of Rayleigh-B\'enard convective cells \cite{King2013} provide scaling 
laws for the viscous boundary layer thickness for non-rotating systems:
\begin{equation}
 \delta\sim 0.34\,Re^{-1/4}\,L,
\end{equation}
where $L$ is the size of the convective cell and $Re$ is the Reynolds number. The 
simulations were performed for systems that are quite different from gas giants, 
with lower Reynolds number than for Jupiter convection for instance ($Re=10^{12}$ 
for the deep interior of Jupiter \cite{Guillot2004}) and higher Prandtl numbers 
(above 1 for the simulations but closer to $10^{-2}$ for Jupiter 
\cite{French2012}). But comparison with experiments in other range of parameters 
give confidence for an extrapolation of this scaling law towards more extreme 
regimes such as the ones in giant planets. If we take $L\sim 10^7$~m for the eddy 
size of the convective cell \cite{Guillot2004}, we find a boundary layer thickness 
of the order of $\delta\sim3$~km. Using the estimate from \fig{fig:diff_heavy} we 
can estimate the diffusion coefficient to be close to 0.1~mm$^2$/s, which results 
in a typical diffusion time scale of $3\times10^6$ years. Such a time scale would 
suggest a nearly complete erosion of the core for a planete of 4.5 billion years. 
But we assumed here a full and efficient convection. Yet, if a semi-convective 
pattern sets in, the efficiency of the mixing could be significantly decreased and 
the erosion process also inhibited. More work in this area is thus necessary.  

\section{\label{sec:conclusion}Conclusion}
Our \textit{ab initio} simulations of H-He-SiO$_2$ mixtures under conditions 
relevant for giant planet interiors showed a smooth dissociation of hydrogen as 
the pressure and the temperature increase. Unlike helium, SiO$_2$ does not 
stabilize molecular hydrogen because oxygen reacts with hydrogen to form water and 
hydroxide molecules. The location of the dissociation and the metallization of 
hydrogen is of first importance because it is related to the demixing of hydrogen 
and helium \cite{Stevenson1977,Morales2009,Lorenzen2011,Soubiran2013}. If hydrogen 
is stabilized or is less metallic, the demixing occurs at higher pressures and, 
thus, deeper in the planet. A more detailed analysis of the electronic properties 
is therefore necessary to identify the effect of the heavy elements on the 
metallization.

The calculation of the diffusion coefficients of the different species shows that 
hydrogen and helium are at most only marginally affected by the presence of the 
heavy elements. We were able to estimate the diffusion coefficients for oxygen and 
silicon in the H-He metallic phase and obtained values between 0.01 and 
1~mm$^2$/s. We also obtained values for the viscosity ranging between 0.1 and 1 
mPa.s indicating a turbulent behavior of the convection in giant planets. These 
ionic transport properties are typically extremely challenging to determine in 
high-pressure experiments, yet crucial for the determination of the dynamics of 
the deep interior of a planet.

Based on our calculated transport properties, we estimate the typical time scale 
for heavy elements to migrate from the core to the envelope, by diffusion through 
the boundary layer, to be of the order of a few million years. This is short on a 
geological time-scale and suggests that the core of giant planets could be 
entirely eroded if the convection is efficient. However, if semi-convection sets 
in, the mixing of the heavy elements into the envelope could be significantly 
slowed down, which would stabilize the core. Additional work is required in this 
area to better characterize the dynamics of the envelope near the core boundary.

\section{Acknowledgments}
We are very thankful to Bruce Buffett for the instructive discussions on the 
convection in planetary interiors. The authors achnowlegde support from the U.S. 
National Science Foundation (AST 1412646) and from the U.S. Department of Energy 
(DE-SC0010517 and DE-SC0016248). Calculations were performed on Pleiades from NAS 
as well as on Comet and Stampede from the XSEDE program.


\begin{thebibliography}{42}%
\makeatletter
\providecommand \@ifxundefined [1]{%
 \@ifx{#1\undefined}
}%
\providecommand \@ifnum [1]{%
 \ifnum #1\expandafter \@firstoftwo
 \else \expandafter \@secondoftwo
 \fi
}%
\providecommand \@ifx [1]{%
 \ifx #1\expandafter \@firstoftwo
 \else \expandafter \@secondoftwo
 \fi
}%
\providecommand \natexlab [1]{#1}%
\providecommand \enquote  [1]{``#1''}%
\providecommand \bibnamefont  [1]{#1}%
\providecommand \bibfnamefont [1]{#1}%
\providecommand \citenamefont [1]{#1}%
\providecommand \href@noop [0]{\@secondoftwo}%
\providecommand \href [0]{\begingroup \@sanitize@url \@href}%
\providecommand \@href[1]{\@@startlink{#1}\@@href}%
\providecommand \@@href[1]{\endgroup#1\@@endlink}%
\providecommand \@sanitize@url [0]{\catcode `\\12\catcode `\$12\catcode
  `\&12\catcode `\#12\catcode `\^12\catcode `\_12\catcode `\%12\relax}%
\providecommand \@@startlink[1]{}%
\providecommand \@@endlink[0]{}%
\providecommand \url  [0]{\begingroup\@sanitize@url \@url }%
\providecommand \@url [1]{\endgroup\@href {#1}{\urlprefix }}%
\providecommand \urlprefix  [0]{URL }%
\providecommand \Eprint [0]{\href }%
\providecommand \doibase [0]{http://dx.doi.org/}%
\providecommand \selectlanguage [0]{\@gobble}%
\providecommand \bibinfo  [0]{\@secondoftwo}%
\providecommand \bibfield  [0]{\@secondoftwo}%
\providecommand \translation [1]{[#1]}%
\providecommand \BibitemOpen [0]{}%
\providecommand \bibitemStop [0]{}%
\providecommand \bibitemNoStop [0]{.\EOS\space}%
\providecommand \EOS [0]{\spacefactor3000\relax}%
\providecommand \BibitemShut  [1]{\csname bibitem#1\endcsname}%
\let\auto@bib@innerbib\@empty
\bibitem [{\citenamefont {Pollack}\ \emph {et~al.}(1996)\citenamefont
  {Pollack}, \citenamefont {Hubickyj}, \citenamefont {Bodenheimer},
  \citenamefont {Lissauer}, \citenamefont {Podolak},\ and\ \citenamefont
  {Greenzweig}}]{Pollack1996a}%
  \BibitemOpen
  \bibfield  {author} {\bibinfo {author} {\bibfnamefont {J.~B.}\ \bibnamefont
  {Pollack}}, \bibinfo {author} {\bibfnamefont {O.}~\bibnamefont {Hubickyj}},
  \bibinfo {author} {\bibfnamefont {P.~H.}\ \bibnamefont {Bodenheimer}},
  \bibinfo {author} {\bibfnamefont {J.~J.}\ \bibnamefont {Lissauer}}, \bibinfo
  {author} {\bibfnamefont {M.}~\bibnamefont {Podolak}}, \ and\ \bibinfo
  {author} {\bibfnamefont {Y.}~\bibnamefont {Greenzweig}},\ }\href {\doibase
  10.1006/icar.1996.0190} {\bibfield  {journal} {\bibinfo  {journal} {Icarus}\
  }\textbf {\bibinfo {volume} {124}},\ \bibinfo {pages} {62} (\bibinfo {year}
  {1996})},\ 
  \BibitemShut {NoStop}%
\bibitem [{\citenamefont {Fortney}\ \emph {et~al.}(2013)\citenamefont
  {Fortney}, \citenamefont {Mordasini}, \citenamefont {Nettelmann},
  \citenamefont {Kempton}, \citenamefont {Greene},\ and\ \citenamefont
  {Zahnle}}]{Fortney2013}%
  \BibitemOpen
  \bibfield  {author} {\bibinfo {author} {\bibfnamefont {J.~J.}\ \bibnamefont
  {Fortney}}, \bibinfo {author} {\bibfnamefont {C.}~\bibnamefont {Mordasini}},
  \bibinfo {author} {\bibfnamefont {N.}~\bibnamefont {Nettelmann}}, \bibinfo
  {author} {\bibfnamefont {E.~M.-R.}\ \bibnamefont {Kempton}}, \bibinfo
  {author} {\bibfnamefont {T.~P.}\ \bibnamefont {Greene}}, \ and\ \bibinfo
  {author} {\bibfnamefont {K.}~\bibnamefont {Zahnle}},\ }\href {\doibase
  10.1088/0004-637X/775/1/80} {\bibfield  {journal} {\bibinfo  {journal}
  {Astrophys. J.}\ }\textbf {\bibinfo {volume} {775}},\ \bibinfo {pages} {80}
  (\bibinfo {year} {2013})}\BibitemShut {NoStop}%
\bibitem [{\citenamefont {Wilson}\ and\ \citenamefont
  {Militzer}(2012{\natexlab{a}})}]{Wilson2012a}%
  \BibitemOpen
  \bibfield  {author} {\bibinfo {author} {\bibfnamefont {H.~F.}\ \bibnamefont
  {Wilson}}\ and\ \bibinfo {author} {\bibfnamefont {B.}~\bibnamefont
  {Militzer}},\ }\href@noop {} {\bibfield  {journal} {\bibinfo  {journal}
  {Astrophys. J.}\ }\textbf {\bibinfo {volume} {745}},\ \bibinfo {pages} {54}
  (\bibinfo {year} {2012}{\natexlab{a}})}\BibitemShut {NoStop}%
\bibitem [{\citenamefont {Wilson}\ and\ \citenamefont
  {Militzer}(2012{\natexlab{b}})}]{Wilson2012b}%
  \BibitemOpen
  \bibfield  {author} {\bibinfo {author} {\bibfnamefont {H.~F.}\ \bibnamefont
  {Wilson}}\ and\ \bibinfo {author} {\bibfnamefont {B.}~\bibnamefont
  {Militzer}},\ }\href@noop {} {\bibfield  {journal} {\bibinfo  {journal}
  {Phys. Rev. Lett.}\ }\textbf {\bibinfo {volume} {108}},\ \bibinfo {pages}
  {111101} (\bibinfo {year} {2012}{\natexlab{b}})}\BibitemShut {NoStop}%
\bibitem [{\citenamefont {Wahl}\ \emph {et~al.}(2013)\citenamefont {Wahl},
  \citenamefont {Wilson},\ and\ \citenamefont {Militzer}}]{Wahl2013}%
  \BibitemOpen
  \bibfield  {author} {\bibinfo {author} {\bibfnamefont {S.~M.}\ \bibnamefont
  {Wahl}}, \bibinfo {author} {\bibfnamefont {H.~F.}\ \bibnamefont {Wilson}}, \
  and\ \bibinfo {author} {\bibfnamefont {B.}~\bibnamefont {Militzer}},\ }\href
  {\doibase 10.1088/0004-637X/773/2/95} {\bibfield  {journal} {\bibinfo
  {journal} {Astrophys. J.}\ }\textbf {\bibinfo {volume} {773}},\ \bibinfo
  {pages} {95} (\bibinfo {year} {2013})}\BibitemShut {NoStop}%
\bibitem [{\citenamefont {Gonz{\'{a}}lez-Cataldo}\ \emph
  {et~al.}(2014)\citenamefont {Gonz{\'{a}}lez-Cataldo}, \citenamefont
  {Wilson},\ and\ \citenamefont {Militzer}}]{Gonzalez-Cataldo2014}%
  \BibitemOpen
  \bibfield  {author} {\bibinfo {author} {\bibfnamefont {F.}~\bibnamefont
  {Gonz{\'{a}}lez-Cataldo}}, \bibinfo {author} {\bibfnamefont {H.~F.}\
  \bibnamefont {Wilson}}, \ and\ \bibinfo {author} {\bibfnamefont
  {B.}~\bibnamefont {Militzer}},\ }\href {\doibase 10.1088/0004-637X/787/1/79}
  {\bibfield  {journal} {\bibinfo  {journal} {Astrophys. J.}\ }\textbf
  {\bibinfo {volume} {787}},\ \bibinfo {pages} {79} (\bibinfo {year}
  {2014})}\BibitemShut {NoStop}%
\bibitem [{\citenamefont {Wong}\ \emph {et~al.}(2004)\citenamefont {Wong},
  \citenamefont {Mahaffy}, \citenamefont {Atreya}, \citenamefont {Niemann},\
  and\ \citenamefont {Owen}}]{Wong2004}%
  \BibitemOpen
  \bibfield  {author} {\bibinfo {author} {\bibfnamefont {M.~H.}\ \bibnamefont
  {Wong}}, \bibinfo {author} {\bibfnamefont {P.~R.}\ \bibnamefont {Mahaffy}},
  \bibinfo {author} {\bibfnamefont {S.~K.}\ \bibnamefont {Atreya}}, \bibinfo
  {author} {\bibfnamefont {H.~B.}\ \bibnamefont {Niemann}}, \ and\ \bibinfo
  {author} {\bibfnamefont {T.~C.}\ \bibnamefont {Owen}},\ }\href {\doibase
  10.1016/j.icarus.2004.04.010} {\bibfield  {journal} {\bibinfo  {journal}
  {Icarus}\ }\textbf {\bibinfo {volume} {171}},\ \bibinfo {pages} {153}
  (\bibinfo {year} {2004})}\BibitemShut {NoStop}%
\bibitem [{\citenamefont {Nettelmann}\ \emph {et~al.}(2012)\citenamefont
  {Nettelmann}, \citenamefont {Becker}, \citenamefont {Holst},\ and\
  \citenamefont {Redmer}}]{Nettelmann2012}%
  \BibitemOpen
  \bibfield  {author} {\bibinfo {author} {\bibfnamefont {N.}~\bibnamefont
  {Nettelmann}}, \bibinfo {author} {\bibfnamefont {A.}~\bibnamefont {Becker}},
  \bibinfo {author} {\bibfnamefont {B.}~\bibnamefont {Holst}}, \ and\ \bibinfo
  {author} {\bibfnamefont {R.}~\bibnamefont {Redmer}},\ }\href {\doibase
  10.1088/0004-637X/750/1/52} {\bibfield  {journal} {\bibinfo  {journal}
  {Astrophys. J.}\ }\textbf {\bibinfo {volume} {750}},\ \bibinfo {pages} {52}
  (\bibinfo {year} {2012})}\BibitemShut {NoStop}%
\bibitem [{\citenamefont {Helled}\ and\ \citenamefont
  {Guillot}(2013)}]{Helled2013}%
  \BibitemOpen
  \bibfield  {author} {\bibinfo {author} {\bibfnamefont {R.}~\bibnamefont
  {Helled}}\ and\ \bibinfo {author} {\bibfnamefont {T.}~\bibnamefont
  {Guillot}},\ }\href {\doibase 10.1088/0004-637X/767/2/113} {\bibfield
  {journal} {\bibinfo  {journal} {Astrophys. J.}\ }\textbf {\bibinfo {volume}
  {767}},\ \bibinfo {pages} {113} (\bibinfo {year} {2013})}\BibitemShut
  {NoStop}%
\bibitem [{\citenamefont {Militzer}\ and\ \citenamefont
  {Hubbard}(2013)}]{Militzer2013b}%
  \BibitemOpen
  \bibfield  {author} {\bibinfo {author} {\bibfnamefont {B.}~\bibnamefont
  {Militzer}}\ and\ \bibinfo {author} {\bibfnamefont {W.~B.}\ \bibnamefont
  {Hubbard}},\ }\href {\doibase 10.1088/0004-637X/774/2/148} {\bibfield
  {journal} {\bibinfo  {journal} {Astrophys. J.}\ }\textbf {\bibinfo {volume}
  {774}},\ \bibinfo {pages} {148} (\bibinfo {year} {2013})},\  \BibitemShut {NoStop}%
\bibitem [{\citenamefont {Hubbard}\ and\ \citenamefont
  {Militzer}(2016)}]{Hubbard2016}%
  \BibitemOpen
  \bibfield  {author} {\bibinfo {author} {\bibfnamefont {W.~B.}\ \bibnamefont
  {Hubbard}}\ and\ \bibinfo {author} {\bibfnamefont {B.}~\bibnamefont
  {Militzer}},\ }\href {\doibase 10.3847/0004-637X/820/1/80} {\bibfield
  {journal} {\bibinfo  {journal} {Astrophys. J.}\ }\textbf {\bibinfo {volume}
  {820}},\ \bibinfo {pages} {80} (\bibinfo {year} {2016})}\BibitemShut
  {NoStop}%
\bibitem [{\citenamefont {Militzer}\ \emph {et~al.}(2016)\citenamefont
  {Militzer}, \citenamefont {Soubiran}, \citenamefont {Wahl},\ and\
  \citenamefont {Hubbard}}]{Militzer2016}%
  \BibitemOpen
  \bibfield  {author} {\bibinfo {author} {\bibfnamefont {B.}~\bibnamefont
  {Militzer}}, \bibinfo {author} {\bibfnamefont {F.}~\bibnamefont {Soubiran}},
  \bibinfo {author} {\bibfnamefont {S.~M.}\ \bibnamefont {Wahl}}, \ and\
  \bibinfo {author} {\bibfnamefont {W.}~\bibnamefont {Hubbard}},\ }\href
  {\doibase 10.1002/2016JE005080.Abstract} {\bibfield  {journal} {\bibinfo
  {journal} {JGR Planets}\ }\textbf {\bibinfo {volume} {121}},\ \bibinfo
  {pages} {1552} (\bibinfo {year} {2016})}\BibitemShut {NoStop}%
\bibitem [{\citenamefont {Militzer}(2013)}]{Militzer2013a}%
  \BibitemOpen
  \bibfield  {author} {\bibinfo {author} {\bibfnamefont {B.}~\bibnamefont
  {Militzer}},\ }\href {\doibase 10.1103/PhysRevB.87.014202} {\bibfield
  {journal} {\bibinfo  {journal} {Phys Rev B}\ }\textbf {\bibinfo {volume}
  {87}},\ \bibinfo {pages} {014202} (\bibinfo {year} {2013})}\BibitemShut
  {NoStop}%
\bibitem [{\citenamefont {Soubiran}\ and\ \citenamefont
  {Militzer}(2016)}]{Soubiran2016}%
  \BibitemOpen
  \bibfield  {author} {\bibinfo {author} {\bibfnamefont {F.}~\bibnamefont
  {Soubiran}}\ and\ \bibinfo {author} {\bibfnamefont {B.}~\bibnamefont
  {Militzer}},\ }\href@noop {} {\bibfield  {journal} {\bibinfo  {journal}
  {Astrophys. J.}\ }\textbf {\bibinfo {volume} {829}},\ \bibinfo {pages} {14}
  (\bibinfo {year} {2016})}\BibitemShut {NoStop}%
\bibitem [{\citenamefont {Stevenson}(1982)}]{Stevenson1982}%
  \BibitemOpen
  \bibfield  {author} {\bibinfo {author} {\bibfnamefont {D.~J.}\ \bibnamefont
  {Stevenson}},\ }\href {\doibase 10.1016/0032-0633(82)90108-8} {\bibfield
  {journal} {\bibinfo  {journal} {Planet. Space Sci.}\ }\textbf {\bibinfo
  {volume} {30}},\ \bibinfo {pages} {755} (\bibinfo {year} {1982})}\BibitemShut
  {NoStop}%
\bibitem [{\citenamefont {Leconte}\ and\ \citenamefont
  {Chabrier}(2012)}]{Leconte2012}%
  \BibitemOpen
  \bibfield  {author} {\bibinfo {author} {\bibfnamefont {J.}~\bibnamefont
  {Leconte}}\ and\ \bibinfo {author} {\bibfnamefont {G.}~\bibnamefont
  {Chabrier}},\ }\href@noop {} {\bibfield  {journal} {\bibinfo  {journal}
  {A{\&}A}\ }\textbf {\bibinfo {volume} {540}},\ \bibinfo {pages} {A20}
  (\bibinfo {year} {2012})}\BibitemShut {NoStop}%
\bibitem [{\citenamefont {Leconte}\ and\ \citenamefont
  {Chabrier}(2013)}]{Leconte2013}%
  \BibitemOpen
  \bibfield  {author} {\bibinfo {author} {\bibfnamefont {J.}~\bibnamefont
  {Leconte}}\ and\ \bibinfo {author} {\bibfnamefont {G.}~\bibnamefont
  {Chabrier}},\ }\href@noop {} {\bibfield  {journal} {\bibinfo  {journal} {Nat.
  Geosci.}\ }\textbf {\bibinfo {volume} {6}},\ \bibinfo {pages} {347} (\bibinfo
  {year} {2013})}\BibitemShut {NoStop}%
\bibitem [{\citenamefont {French}\ \emph {et~al.}(2012)\citenamefont {French},
  \citenamefont {Becker}, \citenamefont {Lorenzen}, \citenamefont {Nettelmann},
  \citenamefont {Bethkenhagen}, \citenamefont {Wicht},\ and\ \citenamefont
  {Redmer}}]{French2012}%
  \BibitemOpen
  \bibfield  {author} {\bibinfo {author} {\bibfnamefont {M.}~\bibnamefont
  {French}}, \bibinfo {author} {\bibfnamefont {A.}~\bibnamefont {Becker}},
  \bibinfo {author} {\bibfnamefont {W.}~\bibnamefont {Lorenzen}}, \bibinfo
  {author} {\bibfnamefont {N.}~\bibnamefont {Nettelmann}}, \bibinfo {author}
  {\bibfnamefont {M.}~\bibnamefont {Bethkenhagen}}, \bibinfo {author}
  {\bibfnamefont {J.}~\bibnamefont {Wicht}}, \ and\ \bibinfo {author}
  {\bibfnamefont {R.}~\bibnamefont {Redmer}},\ }\href {\doibase
  10.1088/0067-0049/202/1/5} {\bibfield  {journal} {\bibinfo  {journal}
  {Astrophys. J. Suppl. Ser.}\ }\textbf {\bibinfo {volume} {202}},\ \bibinfo
  {pages} {5} (\bibinfo {year} {2012})}\BibitemShut {NoStop}%
\bibitem [{\citenamefont {Kresse}\ and\ \citenamefont
  {Furthm{\"{u}}ller}(1996)}]{Kresse1996}%
  \BibitemOpen
  \bibfield  {author} {\bibinfo {author} {\bibfnamefont {G.}~\bibnamefont
  {Kresse}}\ and\ \bibinfo {author} {\bibfnamefont {J.}~\bibnamefont
  {Furthm{\"{u}}ller}},\ }\href {\doibase 10.1103/PhysRevB.54.11169} {\bibfield
   {journal} {\bibinfo  {journal} {Phys. Rev. B}\ }\textbf {\bibinfo {volume}
  {54}},\ \bibinfo {pages} {11169} (\bibinfo {year} {1996})},\  \BibitemShut {NoStop}%
\bibitem [{\citenamefont {Nos{\'{e}}}(1984)}]{Nose1984}%
  \BibitemOpen
  \bibfield  {author} {\bibinfo {author} {\bibfnamefont {S.}~\bibnamefont
  {Nos{\'{e}}}},\ }\href {\doibase 10.1063/1.447334} {\bibfield  {journal}
  {\bibinfo  {journal} {J. Chem. Phys.}\ }\textbf {\bibinfo {volume} {81}},\
  \bibinfo {pages} {511} (\bibinfo {year} {1984})}\BibitemShut {NoStop}%
\bibitem [{\citenamefont {Nos{\'{e}}}(1991)}]{Nose1991}%
  \BibitemOpen
  \bibfield  {author} {\bibinfo {author} {\bibfnamefont {S.}~\bibnamefont
  {Nos{\'{e}}}},\ }\href@noop {} {\bibfield  {journal} {\bibinfo  {journal}
  {Prog. Theor. Phys. Suppl.}\ }\textbf {\bibinfo {volume} {103}},\ \bibinfo
  {pages} {1} (\bibinfo {year} {1991})}\BibitemShut {NoStop}%
\bibitem [{\citenamefont {Kohn}\ and\ \citenamefont {Sham}(1965)}]{Kohn1965}%
  \BibitemOpen
  \bibfield  {author} {\bibinfo {author} {\bibfnamefont {W.}~\bibnamefont
  {Kohn}}\ and\ \bibinfo {author} {\bibfnamefont {L.~J.}\ \bibnamefont
  {Sham}},\ }\href {\doibase http://dx.doi.org/10.1103/PhysRev.140.A1133}
  {\bibfield  {journal} {\bibinfo  {journal} {Phys. Rev.}\ }\textbf {\bibinfo
  {volume} {140}},\ \bibinfo {pages} {1133} (\bibinfo {year}
  {1965})}\BibitemShut {NoStop}%
\bibitem [{\citenamefont {Mermin}(1965)}]{Mermin1965}%
  \BibitemOpen
  \bibfield  {author} {\bibinfo {author} {\bibfnamefont {N.~D.}\ \bibnamefont
  {Mermin}},\ }\href@noop {} {\bibfield  {journal} {\bibinfo  {journal} {Phys.
  Rev.}\ }\textbf {\bibinfo {volume} {137}},\ \bibinfo {pages} {1441} (\bibinfo
  {year} {1965})}\BibitemShut {NoStop}%
\bibitem [{\citenamefont {Perdew}\ \emph {et~al.}(1996)\citenamefont {Perdew},
  \citenamefont {Burke},\ and\ \citenamefont {Ernzerhof}}]{Perdew1996}%
  \BibitemOpen
  \bibfield  {author} {\bibinfo {author} {\bibfnamefont {J.~P.}\ \bibnamefont
  {Perdew}}, \bibinfo {author} {\bibfnamefont {K.}~\bibnamefont {Burke}}, \
  and\ \bibinfo {author} {\bibfnamefont {M.}~\bibnamefont {Ernzerhof}},\
  }\href@noop {} {\bibfield  {journal} {\bibinfo  {journal} {Phys. Rev. Lett.}\
  }\textbf {\bibinfo {volume} {77}},\ \bibinfo {pages} {3865} (\bibinfo {year}
  {1996})}\BibitemShut {NoStop}%
\bibitem [{\citenamefont {Bl{\"{o}}chl}(1994)}]{Blochl1994}%
  \BibitemOpen
  \bibfield  {author} {\bibinfo {author} {\bibfnamefont {P.~E.}\ \bibnamefont
  {Bl{\"{o}}chl}},\ }\href {\doibase 10.1103/PhysRevB.50.17953} {\bibfield
  {journal} {\bibinfo  {journal} {Phys. Rev. B}\ }\textbf {\bibinfo {volume}
  {50}},\ \bibinfo {pages} {17953} (\bibinfo {year} {1994})},\ 
  \BibitemShut {NoStop}%
\bibitem [{\citenamefont {Baldereschi}(1973)}]{Baldereschi1973}%
  \BibitemOpen
  \bibfield  {author} {\bibinfo {author} {\bibfnamefont {A.}~\bibnamefont
  {Baldereschi}},\ }\href {\doibase 10.1103/PhysRevB.7.5212} {\bibfield
  {journal} {\bibinfo  {journal} {Phys. Rev. B}\ }\textbf {\bibinfo {volume}
  {7}},\ \bibinfo {pages} {5212} (\bibinfo {year} {1973})}\BibitemShut
  {NoStop}%
\bibitem [{\citenamefont {Frenkel}\ and\ \citenamefont
  {Smit}(2002)}]{Frenkel2002}%
  \BibitemOpen
  \bibfield  {author} {\bibinfo {author} {\bibfnamefont {D.}~\bibnamefont
  {Frenkel}}\ and\ \bibinfo {author} {\bibfnamefont {B.}~\bibnamefont {Smit}},\
  }\href@noop {} {\emph {\bibinfo {title} {{Understanding Molecular Simulation:
  from Algorithms to Applications}}}}\ (\bibinfo  {publisher} {Academic
  Press},\ \bibinfo {year} {2002})\BibitemShut {NoStop}%
\bibitem [{\citenamefont {Danel}\ \emph {et~al.}(2012)\citenamefont {Danel},
  \citenamefont {Kazandjian},\ and\ \citenamefont {Z{\'{e}}rah}}]{Danel2012}%
  \BibitemOpen
  \bibfield  {author} {\bibinfo {author} {\bibfnamefont {J.~F.}\ \bibnamefont
  {Danel}}, \bibinfo {author} {\bibfnamefont {L.}~\bibnamefont {Kazandjian}}, \
  and\ \bibinfo {author} {\bibfnamefont {G.}~\bibnamefont {Z{\'{e}}rah}},\
  }\href {\doibase 10.1103/PhysRevE.85.066701} {\bibfield  {journal} {\bibinfo
  {journal} {Phys. Rev. E}\ }\textbf
  {\bibinfo {volume} {85}},\ \bibinfo {pages} {066701} (\bibinfo {year}
  {2012})}\BibitemShut {NoStop}%
\bibitem [{\citenamefont {Allen}\ and\ \citenamefont
  {Tildesley}(1987)}]{Allen1987}%
  \BibitemOpen
  \bibfield  {author} {\bibinfo {author} {\bibfnamefont {M.~P.}\ \bibnamefont
  {Allen}}\ and\ \bibinfo {author} {\bibfnamefont {D.~J.}\ \bibnamefont
  {Tildesley}},\ }\href@noop {} {\emph {\bibinfo {title} {{Computer Simulation
  of Liquids}}}}\ (\bibinfo  {publisher} {Clarendon Press},\ \bibinfo {address}
  {Oxford},\ \bibinfo {year} {1987})\BibitemShut {NoStop}%
\bibitem [{\citenamefont {Meyer}\ \emph {et~al.}(2014)\citenamefont {Meyer},
  \citenamefont {Kress}, \citenamefont {Collins},\ and\ \citenamefont
  {Ticknor}}]{Meyer2014}%
  \BibitemOpen
  \bibfield  {author} {\bibinfo {author} {\bibfnamefont {E.~R.}\ \bibnamefont
  {Meyer}}, \bibinfo {author} {\bibfnamefont {J.~D.}\ \bibnamefont {Kress}},
  \bibinfo {author} {\bibfnamefont {L.~A.}\ \bibnamefont {Collins}}, \ and\
  \bibinfo {author} {\bibfnamefont {C.}~\bibnamefont {Ticknor}},\ }\href
  {\doibase 10.1103/PhysRevE.90.043101} {\bibfield  {journal} {\bibinfo
  {journal} {Phys. Rev. E}\ }\textbf {\bibinfo {volume} {90}},\ \bibinfo
  {pages} {043101} (\bibinfo {year} {2014})}\BibitemShut {NoStop}%
\bibitem [{\citenamefont {Collins}\ \emph {et~al.}(2001)\citenamefont
  {Collins}, \citenamefont {Bickham}, \citenamefont {Kress}, \citenamefont
  {Mazevet}, \citenamefont {Lenosky}, \citenamefont {Troullier},\ and\
  \citenamefont {Windl}}]{Collins2001}%
  \BibitemOpen
  \bibfield  {author} {\bibinfo {author} {\bibfnamefont {L.~A.}\ \bibnamefont
  {Collins}}, \bibinfo {author} {\bibfnamefont {S.~R.}\ \bibnamefont
  {Bickham}}, \bibinfo {author} {\bibfnamefont {J.~D.}\ \bibnamefont {Kress}},
  \bibinfo {author} {\bibfnamefont {S.}~\bibnamefont {Mazevet}}, \bibinfo
  {author} {\bibfnamefont {T.~J.}\ \bibnamefont {Lenosky}}, \bibinfo {author}
  {\bibfnamefont {N.~J.}\ \bibnamefont {Troullier}}, \ and\ \bibinfo {author}
  {\bibfnamefont {W.}~\bibnamefont {Windl}},\ }\href {\doibase
  10.1103/PhysRevB.63.184110} {\bibfield  {journal} {\bibinfo  {journal} {Phys.
  Rev. B}\ }\textbf {\bibinfo {volume} {63}},\ \bibinfo {pages} {184110}
  (\bibinfo {year} {2001})}\BibitemShut {NoStop}%
\bibitem [{\citenamefont {Morales}\ \emph {et~al.}(2010)\citenamefont
  {Morales}, \citenamefont {Pierleoni}, \citenamefont {Schwegler},\ and\
  \citenamefont {Ceperley}}]{Morales2010}%
  \BibitemOpen
  \bibfield  {author} {\bibinfo {author} {\bibfnamefont {M.~A.}\ \bibnamefont
  {Morales}}, \bibinfo {author} {\bibfnamefont {C.}~\bibnamefont {Pierleoni}},
  \bibinfo {author} {\bibfnamefont {E.}~\bibnamefont {Schwegler}}, \ and\
  \bibinfo {author} {\bibfnamefont {D.~M.}\ \bibnamefont {Ceperley}},\ }\href
  {\doibase 10.1073/pnas.1007309107} {\bibfield  {journal} {\bibinfo  {journal}
  {Proc. Natl. Acad. Sci. U. S. A.}\ }\textbf {\bibinfo {volume} {107}},\
  \bibinfo {pages} {12799} (\bibinfo {year} {2010})}\BibitemShut {NoStop}%
\bibitem [{\citenamefont {Vorberger}\ \emph {et~al.}(2007)\citenamefont
  {Vorberger}, \citenamefont {Tamblyn}, \citenamefont {Militzer},\ and\
  \citenamefont {Bonev}}]{Vorberger2007}%
  \BibitemOpen
  \bibfield  {author} {\bibinfo {author} {\bibfnamefont {J.}~\bibnamefont
  {Vorberger}}, \bibinfo {author} {\bibfnamefont {I.}~\bibnamefont {Tamblyn}},
  \bibinfo {author} {\bibfnamefont {B.}~\bibnamefont {Militzer}}, \ and\
  \bibinfo {author} {\bibfnamefont {S.~A.}\ \bibnamefont {Bonev}},\ }\href
  {\doibase 10.1103/PhysRevB.75.024206} {\bibfield  {journal} {\bibinfo
  {journal} {Phys. Rev. B - Condens. Matter Mater. Phys.}\ }\textbf {\bibinfo
  {volume} {75}},\ \bibinfo {pages} {024206} (\bibinfo {year} {2007})},\
  \BibitemShut {NoStop}%
\bibitem [{\citenamefont {Soubiran}\ and\ \citenamefont
  {Militzer}(2015{\natexlab{a}})}]{Soubiran2015a}%
  \BibitemOpen
  \bibfield  {author} {\bibinfo {author} {\bibfnamefont {F.}~\bibnamefont
  {Soubiran}}\ and\ \bibinfo {author} {\bibfnamefont {B.}~\bibnamefont
  {Militzer}},\ }\href {\doibase 10.1016/j.hedp.2014.10.005} {\bibfield
  {journal} {\bibinfo  {journal} {High Energy Density Phys.}\ }\textbf
  {\bibinfo {volume} {17}},\ \bibinfo {pages} {157} (\bibinfo {year}
  {2015}{\natexlab{a}})}\BibitemShut {NoStop}%
\bibitem [{\citenamefont {Soubiran}\ and\ \citenamefont
  {Militzer}(2015{\natexlab{b}})}]{Soubiran2015}%
  \BibitemOpen
  \bibfield  {author} {\bibinfo {author} {\bibfnamefont {F.}~\bibnamefont
  {Soubiran}}\ and\ \bibinfo {author} {\bibfnamefont {B.}~\bibnamefont
  {Militzer}},\ }\href {\doibase 10.1088/0004-637X/806/2/228} {\bibfield
  {journal} {\bibinfo  {journal} {Astrophys. J.}\ }\textbf {\bibinfo {volume}
  {806}},\ \bibinfo {pages} {228} (\bibinfo {year}
  {2015}{\natexlab{b}})}\BibitemShut {NoStop}%
\bibitem [{\citenamefont {Caillabet}\ \emph {et~al.}(2011)\citenamefont
  {Caillabet}, \citenamefont {Mazevet},\ and\ \citenamefont
  {Loubeyre}}]{Caillabet2011}%
  \BibitemOpen
  \bibfield  {author} {\bibinfo {author} {\bibfnamefont {L.}~\bibnamefont
  {Caillabet}}, \bibinfo {author} {\bibfnamefont {S.}~\bibnamefont {Mazevet}},
  \ and\ \bibinfo {author} {\bibfnamefont {P.}~\bibnamefont {Loubeyre}},\
  }\href {\doibase 10.1103/PhysRevB.83.094101} {\bibfield  {journal} {\bibinfo
  {journal} {Phys. Rev. B }\ }\textbf {\bibinfo
  {volume} {83}} (\bibinfo {year} {2011}),\
  }\BibitemShut {NoStop}%
\bibitem [{\citenamefont {Guillot}\ \emph {et~al.}(2004)\citenamefont
  {Guillot}, \citenamefont {Stevenson}, \citenamefont {Hubbard},\ and\
  \citenamefont {Saumon}}]{Guillot2004}%
  \BibitemOpen
  \bibfield  {author} {\bibinfo {author} {\bibfnamefont {T.}~\bibnamefont
  {Guillot}}, \bibinfo {author} {\bibfnamefont {D.~J.}\ \bibnamefont
  {Stevenson}}, \bibinfo {author} {\bibfnamefont {W.~B.}\ \bibnamefont
  {Hubbard}}, \ and\ \bibinfo {author} {\bibfnamefont {D.}~\bibnamefont
  {Saumon}},\ }\href
  {http://adsabs.harvard.edu/cgi-bin/nph-data{\_}query?bibcode=2004jpsm.book...35G{\&}link{\_}type=ABSTRACT$\backslash$npapers2://publication/uuid/B26DB9A6-624B-4964-80CD-5BB7E9B570F3}
  {\bibfield  {journal} {\bibinfo  {journal} {In: Jupiter. The planet}\ ,\
  \bibinfo {pages} {35}} (\bibinfo {year} {2004})}\BibitemShut {NoStop}%
\bibitem [{\citenamefont {King}\ \emph {et~al.}(2013)\citenamefont {King},
  \citenamefont {Stellmach},\ and\ \citenamefont {Buffett}}]{King2013}%
  \BibitemOpen
  \bibfield  {author} {\bibinfo {author} {\bibfnamefont {E.~M.}\ \bibnamefont
  {King}}, \bibinfo {author} {\bibfnamefont {S.}~\bibnamefont {Stellmach}}, \
  and\ \bibinfo {author} {\bibfnamefont {B.}~\bibnamefont {Buffett}},\ }\href
  {\doibase 10.1017/jfm.2012.586} {\bibfield  {journal} {\bibinfo  {journal}
  {J. Fluid Mech.}\ }\textbf {\bibinfo {volume} {717}},\ \bibinfo {pages} {449}
  (\bibinfo {year} {2013})}\BibitemShut {NoStop}%
\bibitem [{\citenamefont {Stevenson}\ and\ \citenamefont
  {Salpeter}(1977)}]{Stevenson1977}%
  \BibitemOpen
  \bibfield  {author} {\bibinfo {author} {\bibfnamefont {D.~J.}\ \bibnamefont
  {Stevenson}}\ and\ \bibinfo {author} {\bibfnamefont {E.~E.}\ \bibnamefont
  {Salpeter}},\ }\href {\doibase 10.1086/190479} {\bibfield  {journal}
  {\bibinfo  {journal} {ApJSS}\ }\textbf {\bibinfo {volume} {35}},\ \bibinfo
  {pages} {239} (\bibinfo {year} {1977})}\BibitemShut {NoStop}%
\bibitem [{\citenamefont {Morales}\ \emph {et~al.}(2009)\citenamefont
  {Morales}, \citenamefont {Schwegler}, \citenamefont {Ceperley}, \citenamefont
  {Pierleoni}, \citenamefont {Hamel},\ and\ \citenamefont
  {Caspersen}}]{Morales2009}%
  \BibitemOpen
  \bibfield  {author} {\bibinfo {author} {\bibfnamefont {M.~A.}\ \bibnamefont
  {Morales}}, \bibinfo {author} {\bibfnamefont {E.}~\bibnamefont {Schwegler}},
  \bibinfo {author} {\bibfnamefont {D.}~\bibnamefont {Ceperley}}, \bibinfo
  {author} {\bibfnamefont {C.}~\bibnamefont {Pierleoni}}, \bibinfo {author}
  {\bibfnamefont {S.}~\bibnamefont {Hamel}}, \ and\ \bibinfo {author}
  {\bibfnamefont {K.}~\bibnamefont {Caspersen}},\ }\href@noop {} {\bibfield
  {journal} {\bibinfo  {journal} {PNAS}\ }\textbf {\bibinfo {volume} {106}},\
  \bibinfo {pages} {1324} (\bibinfo {year} {2009})}\BibitemShut {NoStop}%
\bibitem [{\citenamefont {Lorenzen}\ \emph {et~al.}(2011)\citenamefont
  {Lorenzen}, \citenamefont {Holst},\ and\ \citenamefont
  {Redmer}}]{Lorenzen2011}%
  \BibitemOpen
  \bibfield  {author} {\bibinfo {author} {\bibfnamefont {W.}~\bibnamefont
  {Lorenzen}}, \bibinfo {author} {\bibfnamefont {B.}~\bibnamefont {Holst}}, \
  and\ \bibinfo {author} {\bibfnamefont {R.}~\bibnamefont {Redmer}},\ }\href
  {\doibase 10.1103/PhysRevB.84.235109} {\bibfield  {journal} {\bibinfo
  {journal} {Phys Rev B}\ }\textbf {\bibinfo {volume} {84}},\ \bibinfo {pages}
  {235109} (\bibinfo {year} {2011})}\BibitemShut {NoStop}%
\bibitem [{\citenamefont {Soubiran}\ \emph {et~al.}(2013)\citenamefont
  {Soubiran}, \citenamefont {Mazevet}, \citenamefont {Winisdoerffer},\ and\
  \citenamefont {Chabrier}}]{Soubiran2013}%
  \BibitemOpen
  \bibfield  {author} {\bibinfo {author} {\bibfnamefont {F.}~\bibnamefont
  {Soubiran}}, \bibinfo {author} {\bibfnamefont {S.}~\bibnamefont {Mazevet}},
  \bibinfo {author} {\bibfnamefont {C.}~\bibnamefont {Winisdoerffer}}, \ and\
  \bibinfo {author} {\bibfnamefont {G.}~\bibnamefont {Chabrier}},\ }\href
  {\doibase 10.1103/PhysRevB.87.165114} {\bibfield  {journal} {\bibinfo
  {journal} {Phys. Rev. B}\ }\textbf {\bibinfo
  {volume} {87}},\ \bibinfo {pages} {165114} (\bibinfo {year}
  {2013})}\BibitemShut {NoStop}%
\end{thebibliography}
\end{document}